\documentstyle[12pt,oldlfont,aps]{revtex}
\title{The watching operators method in the theory of Frenkel
exciton. Novel criterion of localization and its exact
calculation for the non diagonal disirdered 1D chain's zero-state }
\author{Kozlov G.G.}
\address{Institute of Physics - Dept. Photonica, St-Petersburg,
  e-mail:     gkozlov@photonics.phys.spbu.ru}
\begin{document}

\maketitle

\begin{abstract}

A method is proposed for manipulating with diagrammatic
expansion of Green's function of Frenkel's exciton random
walks on the perfect lattice. The method allows one to
select diagrams, to supply diagrams with factors  containing
information about the number of sites the  diagram has passed through,
etc.  Simple problems  related to the defect lattices are
considered  using the proposed method. The new criterion
of localization of Frenkel exciton - the number  of sites
covered by the wave function -  is established. The number
of sites covered  by the zero state of 1D non-diagonally
disordered  chain is studied. It is shown that this problem
can be solved by calculating the random walks Green's
function with modified diagrammatic expansion.  By means
of the developed  method  an exact  analitical expression
for the number of sites  covered by the zero-state is
obtained  and zero-state is shown to be localized.

\end{abstract}

\section{Introduction}

 Let us consider some lattice with sites $r_1...r_N$ and introduce
matrix ${\bf W}$  with elements

\begin{equation}
W_{rr'}= w(r-r')
\end{equation}

The function $w(r)$ is supposed  known.
The problem of Frenkel exciton random walks along the lattice
$r_1...r_N$ of twolevel atoms,
can be reduced to the matrix of this kind
\cite{Pastur}  and this problem is not the only one.
In this case $w(r-r')$ represents an interaction which is
able to transport an  excitation from atom $r$ to atom $r'$.
The study of  Green's function (GF) which corresponds to
the above matrix (1) is of particular interest \cite{Pastur}.
We write down this function in convenient dimensionless form:

\begin{equation}
 {\bf G}= (1-{\bf W})^{-1}
\end{equation}

If suppose the summation over the repeating indexes,
the following series for GF may be written:

\begin{equation}
G_{lm}=\bigg({1\over 1-{\bf
W}}\bigg)_{lm}=\delta_{lm}+W_{lm}+W_{lr}W_{rm}+ W_{lr}W_{rp}W_{pm}+...
\end{equation}

If we denote matrix element $W_{rp}$ by an arrow directed
from site $r$ to site $p$ , then each product of matrix elements
in (3) is  being represented  by    diagram (trajectory) ,
which  connects sites $l$ and $m$ .  Then (3)
can be rewritten as:

\begin{equation}
G_{lm}=\bigg(\hbox{ 
Sum of all  diagrams, which connect sites $l$ and $m$
} \bigg)
\end{equation}

The similar method has been used in \cite{Fayer}.
The above representation allows one to write down 
the formal  expressions for Green's functions 
of a certain relative problems.
For example, if certain atoms $r_1...r_t$ 
have been removed,  then the expression 
for Green's function must be modified as follows: 

\begin{equation}
G_{lm}=\left(\matrix{\hbox{Sum of all diagrams,  
which connect sites $l$ and $m$,} 
\cr\hbox{ not passing through 
the sites $r_1...r_t$} }\right) 
\end{equation}

If we now deal with lattice whose sites can 
be occupied by atoms with probability $C<1$ 
and we are interested in the averaged Green's 
function, then each diagram $D$ of this 
function should be multiplied by probability 
that all $n(D)$ sites $D$ has passed through, 
are occupied by atoms.  This probability 
is equal to $C^{n(D)}$ and, consequently,  
we can write  for the averaged Green's 
function :

\begin{equation}
\langle G_{lm}\rangle_c=\left(\matrix{\hbox
{Sum of all diagrams $D$, which connect sites $l$ ³ $m$,}
\cr\hbox{each diagram is being multiplied by $C^{n(D)}$,}
\cr\hbox{where $n(D)$ - is the number of sites in the diagram} }\right) 
\end{equation}

In the end if we are interested in $\langle  G_{lm} \rangle_c$, $C\sim
1$, we have to calculate $\partial \langle
G_{lm}\rangle_c/\partial C$ , $C=1$.
Making use of (6), it is easy to obtain:

\begin{equation}
{\partial\langle G_{lm}\rangle_c\over\partial C}
=\left(\matrix{\hbox{
Sum of all diagrams $D$, which connect sites $l$ and $m$,
}\cr\hbox{
Each diagram is being multiplied by $n(D)$, 
}\cr\hbox{where $n(D)$ - is the  number of sites in $D$} }\right) 
\end{equation}

These examples shows that Green's functions, related to various problems,  
can be obtained by manipulating with one and the same series of diagrams.
So, it is desirable to construct some kind of observer who, 
while mov
ing  along the diagram, would  register the sites
which  diagram has visited,  would count  the number of
sites which diagram has passed through and so on.
For example, in order to obtain Green's function (5) the
above observer's  duty should be as follows:
 While moving along the diagram and while visiting one
after another the sites of  this diagram, observer
should check    whether  site he  just has visited
is    one of  $r_1...r_t$ which have been removed  and,
if it is so, observer should multiply  diagram
by zero.

To obtain Green's functions (7) and (6), observer should
count sites which he has passed through and should
multiply the diagram  by  total number of sites $n(D)$
in the diagram or by $C^{n(D)}$ respectively.  In
section 2 we show that in certain cases it is possible
to build up the mathematical realisation of described
observer by replacing matrix elements $W_{rp}$ in Green's
function (4) by specially constructed watching operators.

However, the random walks problems described in section 2
which can be solved up to the very end (at least at present)
by above method is rather simple and are of some interest
only as a demonstration of absence of  errors in the
mathematical scheme of the method.

In section 3 we consider  much more nontrivial
problem related to  the character (localized- delocalized)
of Frenkel's exciton zero-state in 1D disordered chain.
We establish a novel criterion  of localization -
the inverse number of sites covered by the wave
function - and show that the problem of calculation
of this value may be formulated in terms of diagrams
and exactly solved by means of specially built
watching operators.

\section{Watching operators}
\subsection{}

Let us consider the Green's function $G_{0r}$ which is
defined on certain  periodical simple lattice. Let us
now perform the Fourier  transformation with respect
to its second  index -$g_k \equiv \sum_r G_{0r} \exp(\imath kr)$.

\begin{equation}
g_k={1\over 1-f_k},\hbox{ where } f_k\equiv \sum_r w(r)\exp(\imath kr)
\end{equation}

in terms of diagrams:

\begin{equation}
g_k=\left(\matrix{\hbox{Sum of all diagrams
which are starting from the site 0.}
\cr\hbox{Each diagram is being multiplied by
$\exp(\imath kr_f)$, where $r_f$ -}
\cr\hbox{is a coordinate of the last site in the diagram} }\right)
\end{equation}

Let us consider formula (5) and denote by $L$ the set of sites
occupied by atoms and by $T$ the set of sites $r_1...r_t$.

In order to convert function (9) to the Green's function of
this unperfect (defect) system we replace matrix elements
$W_{rp}$  in (9) (which are denoted by diagram's arrows )
by  the operators $\hat W_{rp}$ which act on upon arbitrary
function   $J_q$ ($q$ is an auxiliary variable)  in
accordance with the rule:

\begin{equation}
\hat W_{rp}J_q \equiv {W_{rp}\over N}\sum_{q'}
K(q-q')e^{\imath q'(p-r) }J_{q'},
\hbox{    where   } K(q) \equiv \sum_{R\in L} \exp (\imath qR)
\end{equation}

Summation over $q$ is carried out over the first Brillouin zone.
Now Green's function (9)  becomes  the operator
$g_k\rightarrow \hat g_k$. Let's apply this operator to unit.

Let's consider  a certain diagram and  apply the operator
which  corresponds to the first arrow to unit- $\hat W_{0r} 1$.
It is easy to see that if $r\in L$, then the
result will be $W_{0r}  \exp(\imath qr)$.

Otherwise we obtain zero .
Consequently, if the first arrow of the diagram came to
the empty  site,  the diagram  is being multiplied by zero.
By induction we come to the conclusion that it is a general
 result - if certain arrow of the diagram reaches the empty
site  $\rho$  (i.e. $\rho$ is one of the $r_1...r_t$) the
operator  corresponding to this arrow gives  zero.

We see that operators (10) give the mathematical realisation
of  the observer described in the end of the previous section.
That's  why we will call them  {\it the watching operators}.
Consequently, if we denote  the Green's function corresponding
to the lattice with defects by $d_q$  , the following
formula may be written:

\begin{equation}
\hat g_k 1=\left(\matrix{\hbox{Sum of all diagrams which are
started from site 0,}
\cr\hbox{which are not passed through $r_1...r_t$. }
\cr\hbox{Each diagram is multiplied by $\exp(\imath (k+q)r_f)$, where $r_f$}
\cr\hbox{- the coordinate of the last site of the diagram} }\right) =d_{k+q}
\end{equation}

By replacing $W_{0r}=w(r)$ in the explicit expression
(8) for $g_k$ (9)  by the watching operators (10) and
by setting $k=0$, we obtain the  equation for  $d_q$:

\begin{equation}
1=(1-\hat f_0) d_q, \hbox{    where   } \hat f_0 \equiv
 \sum_r \hat W_{0r}
\end{equation}

The simple calculations shows that $d_q$ may be introduced in the form:

\begin{equation}
d_q={1-\sum\limits_{R\in T} X_R \exp(\imath qR) \over 1-f_q}
\end{equation}

where parameters $X_R$ may be obtained from the equations:

\begin{equation}
\sum\limits_{R'\in T} G_{R-R'} X_{R'}=G_R,
\hbox{ }
\hbox{ }
\hbox{ }
\hbox{ }
R \in T,
\end{equation}

where

\begin{equation}
G_R\equiv {1\over N}\sum_q e^{-\imath q R}/(1-f_q)
\end{equation}

- is the Green's function of the periodical lattice. If the amount of
removed
atoms is not too large, the solution of (14) may be presented  in a
compact form . For  example, if the only one atom have been removed
from the site $R$,  then the solution of (14) is: $X_R=G_R/G_0$.

 The further analysis of (14) is possible either if the empty sites
forms the periodical structure.
It should be pointed out that while obtaining (14) we have supposed
that the site $0\in L$.
If it is not so, then $d_q=0$. Formulas (13) and (14) gives it
(in this case $X_0=1, X_{R\neq0}=0$), but an  exact solution
of  (12) is non-zero and represents the sum of diagrams which
start  from the zero site and  never return back to this
site more.

At last  let us note that the initial site $0$ is not
especial one and  the Green's functions having another
first index  may be calculated  by the same way.

\subsection{}

 In this subsection we will introduce another type of
watching operators  which give possibility to multiply
the diagrams by  factors, containing  the information
about the sites which diagram  has visited.

 Let's introduce the watching operators dependent on a parameter
$\alpha$ which  operate as follows:

\begin{equation}
\hat W_{rp} J_q\equiv W_{rp}\bigg(1+{\alpha-1\over
N}\sum_q\bigg) e^{\imath q(p-r)} J_q
\end{equation}

Let's apply the operator corresponding to certain diagram $D$

containing in  $\hat g_k$ to  some function
$\Phi_q\equiv \sum_R \phi_R\exp (-\imath qR)$
(summation  over all lattice).

Operator corresponding to the first arrow of the
diagram operate as follows: $ \hat
W_{0r} \Phi_q =W_{0r}\exp(\imath qr) \Phi_q'$, where $\Phi_q'=\sum_R
\phi_R'\exp(-\imath qR)$,  and $\phi_R'=\phi_R$ if $R \neq r$
and $\phi_r'=\alpha \phi_r$.

  Moving further along the diagram and  applying step by
step the watching  operators corresponding to the  arrows
of the diagram , one can see that the diagram $D$ will
be multiplied by
$\exp(\imath qr_f(D))\tilde\Phi_q(D)=
\exp(\imath r_f(D))\sum_R \tilde\phi_R(D)\exp
 (-\imath qR)$, and if $R\ne0$,
  $\tilde\phi_R(D)=\phi_R \alpha^{(\hbox{\small{The number of visits
of the site R by the diagram D}})}$. $r_f(D)$ - is the last
site of the diagram $D$.

The site 0 ( which the diagram started from) is being taken
into account in a special way:
$\tilde\phi_0(D)=\phi_0 \alpha^{(\hbox
{\small{number of diagram's D visits of site 0 -1}})}$.
 Consequently, $\tilde\Phi_q(D)$ carry the information
about sites, which the  diagram $D$ has visited  and about
corresponding number of visits.
For this reason we call this function {\it carrier}.
The arbitrary function $\Phi_q$, which  we have acted on
upon by operator, may be called
the {\it carrier in the initial state}.
So, we see, that after applying $\hat g_k$ (containing the
watching  operators (16)) to the carrier in initial state
$\Phi_q$, each  diagram is being multiplied  by her own carrier.
Let's now obtain the function (7) by means of the watching
operators (16).
 It is clear now, that if we take the carrier in initial state
in the form: $\Phi_q=\sum_{R\neq 0} \exp(-\imath qR)+\alpha=
N\delta_{q0}+\alpha-1$
and apply $\hat g_k$ (containing the watching operators (16)),
the  following formula is justified:

\begin{equation}
\hat g_k \Phi_q=\left(\matrix{
\hbox{Sum of all diagrams which are started from site 0.}
\cr\hbox{Each diagram is being multiplied by carrier:
 $\sum_R e^{-\imath qR}\alpha^{V(D,R)}$,}
\cr\hbox{Where $V(D,R)$ - the number of diagram's $D$ visits of site $R$.}
\cr\hbox{Each diagram is being multiplied by $e^{\imath (k+q)r_f}$,}
\cr\hbox{where $r_f$ - is the coordinate of the last site of the
diagram.} }\right)
\equiv \gamma_q
\end{equation}

The equation for $\gamma_q$:  $\Phi_q=(1-\hat f_k)\gamma_q$ ( being built
by using the watching operators (16)) may be easily solved in terms of
Green's functions (8) and (15):

\begin{equation}
\gamma_q=Ng_k\delta_{q0}+{(\alpha-1)g_{q+k}g_k\over
1-(\alpha-1)(G_0-1)}
\end{equation}

The comparison of the diagrammatic formulas (7) and (17) shows, that
the Fourier transformation of  (7) with respect to the second index
(similar to that in (8)) may be represented in the form:

\begin{equation}
{\partial \langle g_k\rangle_c\over\partial C} (C=1)=
N g_k-\gamma_0(\alpha=0)={g_k^2\over G_0}
\end{equation}

An other types of the watching operators are also possible \cite{Kozlov}.
In the next section we apply the developed method for solving the
problem which  considerably differs from those, which we have
considered in this section.

\section{The exploration of Frenkel exciton zero-state by
means  of the watching operators method}
\subsection{The criterion of localisation}

Let us consider the random process on the 1D chain of
sites $i=1,2,3...n$, which is defined by the following recurrent relation:

\begin{equation}
\varphi_i=\eta_i \varphi_{i-1},\varphi_0=1
\end{equation}

Where $\eta_i\equiv U_i/U'_i$, and $U_i$ , $U'_i$ - are
independent, similarly distributed,positive random variables,
which can have the values $a>0$ and $b>0$ with
probabilities $C$ and $1-C$ respectively
($a>b$).
$\varphi_i$ represents the square module of the
Frenkel's exciton zero-state in the 1D non-diagonal
disordered chain \cite{Theodorou}, when the Hamiltonian matrix
elements can have values $\sqrt a$ and $\sqrt b$
with probabilities $C$ and $1-C$ respectively.
Let us characterize the zero-state by the number
of sites $N^*$, covered by the corresponding wave
function and calculate this number as follows.
 If $\varphi_i=0$ on certain site $i$  \footnote{In the considered
random process always  $\varphi_i \neq 0$, but it is not important
for this reasoning } , then this site is not covered and is being
not taken into account while calculating $N^*$.
Vice versa, if $\varphi_i$ reaches its maximum $\varphi_{max}$ on
the site  $i$, then this site is covered completely and its
contribution while calculating $N^*$ is equal to unit.
In the general case, the contribution of the arbitrary site $i$ is equal
to $\varphi_i/\varphi_{max}$ and we obtain the following formula:

\begin{equation}
N^*={\sum_i \varphi_i\over \varphi_{max}}
\end{equation}

$1/N^*$ may be treated as a criteria of localisation - it is
finite when zero-state is localized and is going down to
zero as the inverse chain length $n^{-1}$,  if zero-state is delocalized.

 If $\lim\limits_{n\rightarrow \infty} N^* < \infty $,
then for the averaged quantities the following formula
may be written:

\begin{equation}
\lim\limits_{n\rightarrow\infty}\langle N^*\rangle =
\langle\langle N^*\rangle\rangle
\end{equation}

Where $\langle\rangle$  denotes the averaging over the
realisations of the random process and $\langle\langle\rangle\rangle$
denotes the averaging over all chains lengths $n$.
 In the next subsection we will describe the method for exact
calculation of $\langle\langle N^* \rangle\rangle$  by means
of methods developed in section 2.

\subsection{The diagrammatic representation of the random
process $\varphi_i$.  The method for calculating
$\langle \langle N^* \rangle \rangle $}

Let us introduce the diagrammatic representation of
random process $\varphi_i$ as follows.
The four possible values are acceptable for $\varphi_i$.
Each value may be represented by horisontal and vertical arrows as follows:

\begin{equation}
\eta_i=\cases{
a/a\hbox{                       } \uparrow\hbox{   =   }C^2t\cr
b/b\hbox{                       }\downarrow\hbox{   =   }(1-C)^2t\cr
a/b\hbox{                       }\rightarrow\hbox{   =   }(1-C)Ct\cr
b/a\hbox{                       }\leftarrow\hbox{   =   }(1-C)Ct
}
\end{equation}

The length of each arrow is supposed to be equal to unit.
Then the arbitrary realisation of random process $\varphi_i$,
related to the chain consisting of $n$ sites, is being
represented by a trajectory - diagram - on the two dimensional
square lattice. These diagrams has  $n$ arrows. Without loss
of generality we can suppose that each diagram  is being
started from the zero site of this lattice.

Now let us ascribe to the $i$-th arrow of the diagram
the probability of the corresponding value of $\eta_i$,
in the same manner it have been done in (23).
 The parameter $t$ , $0<t<1$ -is an auxiliary one
and in the end of the calculation $t\rightarrow 1$.

Consequently, the diagram   completely describes
the distribution of $\eta_i$ in corresponding
realisation and the diagram's value is equal to the
probability of this realisation.
Let's now explain how the $\varphi_{max}$
and $\sum_i \varphi_i $ can be extracted from the diagram.

Introduce the coordinate system $XY$ with zero point
coinciding  with the zero of the above mentioned square lattice.

If $r$ is the $x$- projection of the right
extremal site (sites) of the diagram,
then $\varphi_{max}=\xi^{r}$, where $\xi\equiv a/b>1$.

In order to calculate $\sum_i\varphi_i$  for
certain realisation, let us suppose that we
have multiplied the corresponding diagram
$D$ by the following carrier
$\tilde\Phi_q(D)=\sum_xV(D,x)\exp (-\imath qx)$,
where $V(D,x)$ - the number of diagram's visits of the
sites whose {\it $x$- projection}
(contrary to (17)) is equal to $x$.

After the Fourier transformation the
carrier becomes: $\phi_x(D)=V(D,x)$.
Then $\sum_i\varphi_i = \sum_x\xi^x\phi_x$.

Now let's introduce the matrix ${\bf W}$ (1)
defined on the above mentioned square lattice
by means of the following function $w({\bf r})$:

\begin{equation}
w({\bf r})=\cases{
C^2t,\hbox{ if  } {\bf r}=(0,1)\hbox{             }\uparrow\cr
(1-C)^2t,\hbox{ if  } {\bf r}=(0,-1)\hbox{           }\downarrow\cr
(1-C)Ct,\hbox{ if  } {\bf r}=(1,0)\hbox{           }\rightarrow\cr
(1-C)Ct,\hbox{ if  } {\bf r}=(-1,0)\hbox{          }\leftarrow
},
\end{equation}
and build up the following function in accordance
with the first section:

\begin{equation}
F(q,q',k,\alpha,\beta)=\left(\matrix{
\hbox{Sum of all diagrams which are starting from the site 0.}
\cr\hbox{Each diagram is being multiplied by the
following carrier:}
\cr\hbox{$\bigg(\sum_x e^{-\imath qx}\alpha^{V(D,x)}\bigg)
\bigg(\sum_x e^{-\imath q'x}\beta^{V(D,x)}\bigg)$,}
\cr\hbox{where $V(D,x)$ - is a number of diagram's $D$ visits}
\cr\hbox{of the sites whose  $x$- coordinate is $x$.}
\cr\hbox{The diagram is also multiplied by }
\cr\hbox{$e^{\imath (k+q+q')x_f}$, where $x_f$ - $x$- coordinate}
\cr\hbox{ of the last site of the diagram} }\right)
\end{equation}

Defining the  operation $D/Dx$ as $Dy/Dx \equiv y(x+1)-y(x) $,
we introduce the function:

\begin{equation}
U(x,x')\equiv
-{D\over Dx}\bigg({1\over N}\bigg)^2\sum\limits_{qq'}e^{\imath(qx+q'x')}
\int\limits_0^1 d\alpha {\partial\over\partial\alpha}
 {\partial\over\partial\beta}F(q,q',-q-q',\alpha,1)
\end{equation}

Where $N$ - is the number of sites along the $x$- direction.
$N \rightarrow \infty$ in the end of the calculations.
In terms of diagrams the expression for $U(x,x')$ is as follows:

\begin{equation}
U(x,x')=\left(\matrix{
\hbox{Sum of all diagrams which are being started from}
\cr\hbox{the site 0. All diagrams are being multiplied by the carrier}
\cr\hbox{$\bigg(\delta_{x,r}-\delta_{x,l-1} \bigg)
 V(D,x')$,}
\cr\hbox{where $V(D,x')$ - is the number of diagram's $D$ visits }
\cr\hbox{of the sites whose $x$- coordinate is $x$. }
\cr\hbox{Where $r$($l$) - $x$- projection of the right (left) extremal site}
\cr\hbox{ (or sites) of the diagram $D$.}
 }\right)
\end{equation}

Let us explain the sense of operation $D/Dx$ in (26).
After the Fourier transformation with respect to $q$ and integration
over $\alpha$, the dependence on $x$ for carrier of any diagram is
being represented by a set of {\it ordered} $\delta$- symbols:
$\delta_{x,r}+\delta_{x,r-1}+\delta_{x,r-2}+...\delta_{x,l}$,
where $r$($l$) - is the $x$- projection of the right (left) extremal
site (or sites) of the diagram.

The absence of gaps in the series of the $\delta$- symbols is
take place because an arrows, which  correspond to the
function $w({\bf r})$ (24),   can connect only  neighbour
sites (along the $x$- direction)

Consequently, after operation $D/Dx$ only right $\delta_{x,r}$
and left $\delta_{x,l-1}$  $\delta$- symbols (shifted per unit)
are survive. Then the following formula for the average number
of sites (22) is justified:

\begin{equation}
\langle\langle N^*\rangle\rangle=
\lim\limits_{t\rightarrow1}(1-t)\sum\limits_{x=0}^{\infty}
\sum\limits_{x'=-\infty}^{\infty}
U(x,x')\xi^{x'-x}
\end{equation}

The $1-t$ - is the normalisation factor  related to the averaging
over chain lengths and is equal to inverse sum of all diagrams,
i.e. $g_0^{-1}$  (formula (8), constructed by means of (24)).
The explicit expression for the function (25) can be obtained
by manipulating with   diagrammatic expansion of the Green's
functions (8),(9), by means of the watching operators of the
following form:

\begin{equation}
\hat w({\bf r})=\cases{
 {\bf r}=(0,1) \hbox{       } \hbox{       }
 C^2 t \big[1+{\alpha-1\over N}\sum_q\big]
 \big[1+{\beta-1\over N}\sum_{q'}\big]
\cr
 {\bf r}=(0,-1) \hbox{      } \hbox{       }
 (1-C)^2t \big[1+{\alpha-1\over N}\sum_q\big]
 \big[1+{\beta-1\over N}\sum_{q'}\big]
\cr
 {\bf r}=(1,0) \hbox{       } \hbox{       }
 (1-C)Ct \big[1+{\alpha-1\over N}\sum_q\big]
 \big[1+{\beta-1\over N}\sum_{q'}\big] e^{\imath (q+q')}
\cr
{\bf r}=(-1,0)\hbox{       } \hbox{       }
 (1-C)Ct\big[1+{\alpha-1\over N}\sum_q\big]
 \big[1+{\beta-1\over N}\sum_{q'}\big]e^{-\imath (q+q')}
},
\end{equation}

These operators differ from (16) by  that depend on two
auxiliary wave vectors $q,q'(-\pi<q,q'<\pi)$ , and that
they watch only $x$ projections of the diagrams arrows on
two dimensional lattice.
By applying   $\hat g_k$ to $(N\delta_{q0}+\alpha-1)
(N\delta_{q'0}+\beta-1)$, we obtain the standart
equation for $F(q,q',k,\alpha,\beta)$:

\begin{equation}
(1-\hat f_k)F=(N\delta_{q0}+\alpha-1)(N\delta_{q'0}+\beta-1)
\end{equation}
The solution of this equation is rather clumsy. It may be
reduced to the integral equation of convolution type, which
can be solved by Fourier transformation.
The solution can be expressed in terms of known Green's
function of the Frenkel's exciton on one dimensional
lattice with only nearest neighbour interaction.

Using (25),(26),(28) we obtain the exact expression for $N^*$:

\begin{equation}
\langle\langle N^* \rangle\rangle={\xi(\xi+1)\over (\xi-1)^2C(1-C)},
\hbox{          }
\hbox{          }
\hbox{          }
\hbox{          }
\xi>1
\end{equation}

Let us to note in the conclusion that the above  problem is not
the most simple problem, related to the random process (20),
which can be solved by means of the watching operators method  .
By applying the diagrammatic representation of the random
process (20) and the watching operators method one would
obtain such nontrivial characteristics of (20) as
distribution function for $\varphi_{max}$ and the dependence
of $\langle \varphi_{max} \rangle$ as a function of  chain's
length $n$.

We present the result (31) as an argument in favour of localized
type of the zero state, contrary to \cite{Theodorou}, where zero
state was declared to be delocalized because of its  zero inverse
localization length (ILL).
This criterion of localization is   not correct because the
above model also has zero ILL, but (31) shows that zero
state is always localized.

\end{document}